\documentclass[preprint2]{aastex}

\slugcomment{submitted to APJL.}

\shorttitle{Cm-grains in Oph IRS 48}
\shortauthors{van der Marel et al.}

\begin{document}

\title{A concentration of centimeter-sized grains in the Ophiuchus IRS 48 dust trap}


\author{N. van der Marel, P. Pinilla, J. Tobin\altaffilmark{1} and T. van Kempen}
\affil{Leiden Observatory, P.O. Box 9513, 2300 RA Leiden, the Netherlands}
\email{nmarel@strw.leidenuniv.nl}

\and
\author{S. Andrews, L. Ricci and T. Birnstiel}
\affil{Harvard-Smithsonian Center for Astrophysics, 60 Garden Street, Cambridge, MA 02138}

\altaffiltext{1}{VENI fellow}

\begin{abstract}
Azimuthally asymmetric dust distributions observed with the Atacama Large Millimeter/submillimeter Array (ALMA) in transition disks have been interpreted as dust traps. We present Very Large Array (VLA) Ka band (34 GHz or 0.9 cm) and ALMA Cycle 2 Band 9 (680 GHz or 0.45 mm) observations at a 0$\farcs$2 resolution of the Oph IRS 48 disk, which suggest that larger particles could be more azimuthally concentrated than smaller dust grains, assuming an axisymmetric temperature field or optically thin 680 GHz emission. Fitting an intensity model to both data demonstrates that the azimuthal extent of the millimeter emission is 2.3 $\pm0.9$ times as wide as the centimeter emission, marginally consistent with the particle trapping mechanism under the above assumptions. The 34 GHz continuum image also reveals evidence for ionized gas emission from the star. Both the morphology and the spectral index variations are consistent with an increase of large particles in the center of the trap, but uncertainties remain due to the continuum optical depth at 680 GHz. Particle trapping has been proposed in planet formation models to allow dust particles to grow beyond millimeter sizes in the outer regions of protoplanetary disks. The new observations in the Oph IRS 48 disk provide support for the dust trapping mechanism for centimeter-sized grains, although additional data are required for definitive confirmation.
\end{abstract}



\section{Introduction}
Studies of transitional disks -- protoplanetary disks with inner holes in their dust distribution -- are revolutionizing our understanding of planet formation
(see the review by \citealt{Espaillat2014}). High spatial resolution Atacama Large Millimeter/submillimeter Array (ALMA) observations reveal not only the dust cavities, but also azimuthal asymmetries  \citep{vanderMarel2013,Casassus2013,Fukagawa2013,Perez2014}, and gas still present inside the dust cavities \citep{vanderMarel2013,Casassus2013,Bruderer2014,Zhang2014,vanderMarel2015-12co,SPerez2015}. 
Dust trapping has been suggested as a solution for the radial drift problem \citep{Whipple1972, Weidenschilling1977, Brauer2008} which prevents dust grains from growing beyond millimeter-sizes in the outer disk. Trapping facilitates the crucial steps in dust growth toward the formation of planetesimals (and therefore planets). Because dust continuum emission is dominated by grains with sizes up to three times the observing wavelength \citep{Draine2006}, different continuum wavelengths probe different particle sizes. Since particle trapping depends on particle size, disk turbulence, and the pressure gradient profile, multi-wavelength observations are required to confirm the trapping scenario and constrain disk parameters such as viscosity \citep[e.g.,][and references therein]{Birnstiel2013}.

The Oph IRS 48 transition disk exhibits a highly asymmetric structure in the 0.45 mm dust continuum. The continuum emission is generated by millimeter-sized grains gathered in a peanut-shaped structure, spanning less than a third of the disk ring azimuth; the peak of this structure is $>100\times$ higher than the opposite side of the ring \citep{vanderMarel2013}. However, the gas traced by $^{12}$CO $J$=6--5 line emission has an axisymmetric disk distribution down to 20 AU in radius \citep{Bruderer2014}, confirmed by CO isotopologue observations of the same disk (van der Marel et al. submitted). Likewise, the thermal mid-infrared and near-infrared scattered light emission, tracing small micrometer-sized grains, suggest a ring-like structure although variations along the ring cannot be seen due to high optical depth \citep{Geers2007,Follette2015}. A separation between large and small grains/gas indicates trapping of the large grains in a pressure trap \citep{BargeSommeria1995,KlahrHenning1997,Brauer2008,Pinilla2012a}. Azimuthal traps may result from vortices, which can be due to instabilities such as the Rossby wave or baroclinic instability \citep[e.g.,][]{LyraLin2013,Raettig2013,Fung2014,Flock2015}. A vortex locally increases the gas pressure over a limited radial and azimuthal extent. In the outer disk, millimeter-sized particles will then drift toward this pressure maximum and get trapped in the azimuthal direction \citep[e.g.,][]{Ataiee2013,Zhu2014}, which can explain the observed features in IRS 48. Inside dust traps, the dust particles continue to grow to the maximum grain size permitted by the fragmentation barrier \citep{Birnstiel2010}, which in the outer disk corresponds to centimeter-sizes \citep[e.g.,][]{Pinilla2012a}. Further growth to planetesimal sizes can occur by streaming instabilities \citep{Johansen2009} or by taking mass transfer effects into account \citep{Windmark2012}.

The trapping depends on the particle size: small dust particles are strongly coupled to the gas and thus follow the gas distribution, while larger particles (traced at longer wavelengths) are less coupled and therefore feel the drag force toward the pressure maximum, resulting in a more spatially concentrated distribution of these particles, both radially and azimuthally. Both the trapping concentration and the maximum particle size in the trap are dependent on the turbulence and the gas surface density \citep{Pinilla2015}. In this paper, we present spatially resolved observations of Oph IRS 48 at both millimeter and centimeter wavelengths, taken with ALMA and the Karl G. Jansky Very Large Array (VLA). The distributions of dust emission at 450 $\mu$m and 0.9 cm are compared and we aim to provide confirmation of the dust trap scenario in a transitional disk. Section 2 describes the observation setup and calibration process, Section 3 presents the images and the derived intensity profile and Section 4 discusses the interpretation of the different morphologies and possible implications for the dust trapping scenario. 

\section{Observations}
\begin{table*}[!ht]
\caption{Image properties}
\label{tbl:obsproperties}
\begin{tabular}{llllllll}
\hline
Telescope&Frequency&Bandwidth&Beam size&Beam PA&Flux&Peak&rms\\
&(GHz)&(GHz)&&&(mJy)&(mJy beam$^{-1}$)&(mJy beam$^{-1}$)\\
\hline
ALMA&680&4.7&0$\farcs$19$\times$0$\farcs$15&79$^{\circ}$&1000&190&0.6\\ 
VLA&34&8.0&0$\farcs$46$\times$0$\farcs$26&21$^{\circ}$&251$\times10^{-3}$&138$\times10^{-3}$&3.5$\times10^{-3}$\\ 
\hline
\end{tabular}
\end{table*}

\begin{figure*}[!ht]
\begin{center}
\epsscale{2.2}
\plotone{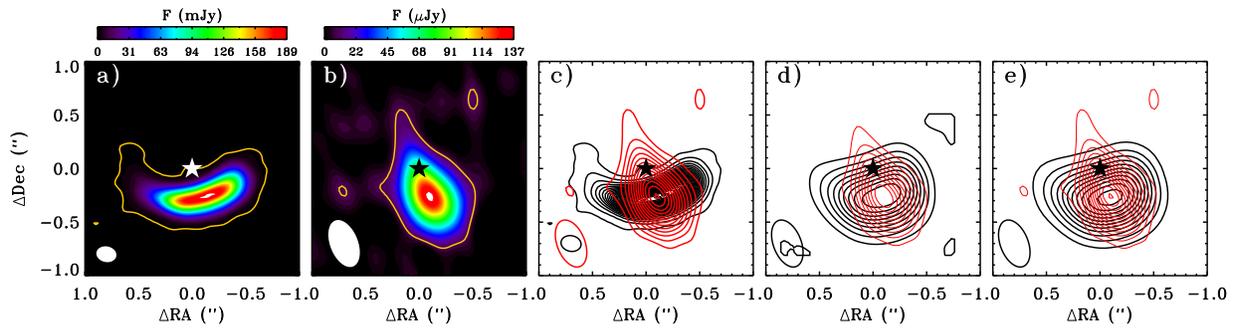} 
\end{center}
\caption{ALMA and VLA observations of dust emission at 680 and 34 GHz of the Oph IRS 48 disk. (a) The 680 GHz image (color scale); (b) the 34 GHz (color scale); (c) the overlay of the 34 GHz contours in red (taken at 3,6,...,39$\sigma$, $\sigma$=3.5 $\mu$Jy) on the 680 GHz contours in black (taken at 3,12, 24,...,324$\sigma$, with $\sigma$=0.6mJy); (d) the overlay of the observed 34 GHz image (red) on the 680 GHz image imaged with the 34 GHz beam (black) in contours taken at 10\%,20\%,...,90\% of the peak; and (e) the overlay of the 680 GHz image as if observed by the VLA (sampled on the VLA visibilities) with the 34 GHz observations taken at 3,6,9,12,...$\sigma$, with $\sigma=3.5\mu$Jy. Ellipses indicate the beam size listed in Table \ref{tbl:obsproperties}. 
\label{fig:dataimage}}
\end{figure*}

Observations of Oph IRS 48 at 34 GHz (9 mm) were obtained using the VLA in the CnB and B configurations in January-February 2015 as part of program 14B-115, with baselines ranging from 75 to 8700 m. The spectral windows were configured for a maximum possible continuum bandwidth of 8 GHz centered at 34 GHz (Ka band) in dual polarization, using 64 spectral windows of 128 MHz each, with 3-bit sampling. Due to the low declination, the source was observed in three scheduling blocks of 2.75 hours with 1.25 hours on source in each block. The bandpass was calibrated using J1517-2422 (in the third block J1924-2914 was used instead), the absolute flux was calibrated using 3C286 and J1625-2527 was used as gain calibrator, periodically observed every three minutes. The pointing was checked on the gain calibrator in X-band every 30 minutes. The calibrated data were concatenated and imaged using Briggs weighting with a robust parameter of 0.5. Deconvolution using CASA imfit reveals that the source is marginally resolved. The flux calibration uncertainty is 10\%.

Observations at 680 GHz (440 $\mu$m) were obtained using ALMA in Cycle 2 in July 2014 in the C34-3 configuration with Band 9 \citep{BaryshevBand9} as part of program 2013.1.00100.S. The observations were taken in four spectral windows of 1920 channels: three windows have a bandwidth of 937.5 MHz, centered on 661, 659 and 675 GHz, the fourth spectral window was centered on 672 GHz with a bandwidth of 1875 MHz. The total continuum bandwidth was $\sim$4.7 GHz. 
The flux was calibrated using J1517-243, the bandpass with J1427-4206 and the phase with J1626-2951. The total on-source integration time was 52 minutes. The data were self-calibrated and imaged using Briggs weighting with a robust parameter of 0.5. The flux calibration uncertainty is 20\%. 

Table \ref{tbl:obsproperties} lists the properties of the images. The astrometric accuracy is set by the calibrators, and is typically $\precsim$30 mas for these two data sets, which is much smaller than the beam size. 

\section{Results}
Figure \ref{fig:dataimage} shows images of the dust continuum emission at 680 and 34 GHz. The 680 GHz continuum has a similar asymmetry as observed in \citet{vanderMarel2013} in the Band 9 Cycle 0 data, though these new data have improved angular resolution. The 34 GHz continuum emission peaks at the same location, but is much more azimuthally concentrated. This is not a sensitivity effect: when the 680 GHz data are restored with the same cleaning beam as the 34 GHz data, its azimuthal extent is still clearly wider than in the measured 34 GHz data (see Figure \ref{fig:dataimage}d). The radial width cannot be compared due to the vertically elongated VLA beam shape, which is caused by the low declination of the source with respect to the VLA site, but given the disk geometry, this elongation does not affect constraints on the azimuthal width, the key parameter of interest here. Spatial filtering can be ruled out as an explanation for the different azimuthal extents; these data recover all of the flux found on the shortest baselines from previous VLA observations in the DnC configuration (beam size 3.3$\times$1.3'', flux = 252 $\pm11 \mu$Jy). Also, we have simulated the ALMA image as if observed by the VLA by sampling the ALMA image on the VLA visibilities (Figure \ref{fig:dataimage}e) to rule out horizontal spatial filtering. Besides, the largest angular scales recovered by the CnB observation is 5", which is much larger than the disk. Thus, the centimeter-sized dust grains (traced by the 34 GHz continuum) have a narrower azimuthal distribution than the millimeter-sized dust grains (traced by the 690 GHz continuum). The peak brightness temperatures for the 680 GHz and 34 GHz are 31 and 1 K, respectively. 

\begin{figure}[!ht]
\begin{center}
\epsscale{1.}
\plotone{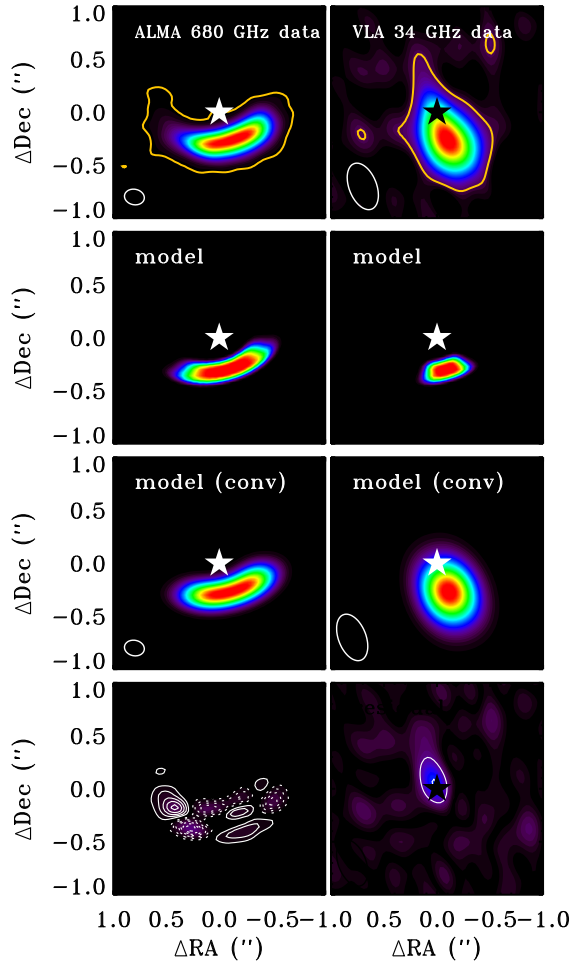}
\end{center}
\caption{Best fits of the intensity profile given in Equation (1) to the dust continuum for the 680 GHz (left) and 34 GHz (right) data. The top panel shows the data in colorscale (the yellow contour shows the 3$\sigma$ level), the middle two panels the unconvolved and convolved model, and the bottom panel the residual image, with the absolute values of the residual in the same color scale as the data image, and overlaid contours at 5$\sigma$ spacing ($\sigma_{\rm 680GHz}$=0.6 mJy beam$^{-1}$, $\sigma_{\rm 34GHz}$=3.5 $\mu$Jy beam$^{-1}$). Dashed contours indicate negative intensity levels. 
\label{fig:bestfit}}
\end{figure}

In order to quantify the concentration of the dust grains, the morphology of the continuum image is fit to a two-dimensional fourth power Gaussian intensity profile $I(r,\phi)$ \citep{vanderMarel2013}:
\begin{equation}
I(r,\phi) = I_c\exp\left(\frac{-(r-r_c)^4}{2r_w^4}\right) \exp\left(\frac{-(\phi-\phi_c)^4}{2\phi_w^4}\right)
\label{eqn:int}
\end{equation}
Previous studies describe azimuthal dust asymmetries due to vortices as regular second power Gaussians \citep{LyraLin2013,Perez2014}, but it was found that our data are fitted much better with the 4$^{th}$ power equation. The stellar position was set to 16$^h$27$^m$37$^s$.185 -24$^{\circ}$30'35$\farcs$39, taken from a Keplerian model fit to the CO isotopologue data from the same ALMA data set (van der Marel et al.\ submitted). The proper motion is negligible compared to the beam size for the period in between these observation sets ($\sim$6 months), so they can be overlaid directly. For the fitting, we use $\chi^2$ minimization over a grid of $\phi_c,\phi_w,r_c$ and $r_w$ with steps of 1$^{\circ}$ and 1 AU for angle and radius respectively. The fitting is performed in the $uv$-plane, with the model visibilities sampled onto the observed spatial frequencies. We find the best fit (see Figure \ref{fig:bestfit}) for $r_c$=61($\pm$ 2) AU, $r_w$=14($\pm$2) AU, $\phi_c$=100($\pm$3)$^{\circ}$ with $\phi_w$=41($\pm$4)$^{\circ}$ for the 680 GHz and $\phi_w$=18($\pm$7)$^{\circ}$ for the 34 GHz continuum. The area at 680 GHz is smaller than that reported in \citet{vanderMarel2013}, due to better spatial resolution, although still within the derived error bars. The values and errors for $r_c$, $r_w$ and $\phi_c$ are mainly constrained by the ALMA data, as the VLA data did not have sufficient spatial resolution to constrain the error bars. The errors are estimated by rescaling the reduced $\chi^2$ to 1. 

The important conclusion is that the azimuthal width of the centimeter emission is 2.3($\pm$0.9) times narrower than the millimeter emission. The aspect ratio of the submillimeter emission is 3.1($\pm$0.6), and $>$1.4 for the centimeter emission. 

The residual of the 680 GHz image still shows significant emission outside the Gaussian fit (peak 25$\sigma$), especially at the tail in the east: 15\% of the total absolute flux remains in the residual. The intensity equation described by Eqn. \ref{eqn:int} is clearly not sufficient to describe the detailed structure: additional (vertical) features may be present. The disk is known to have a large scale height \citep{Bruderer2014}, which is not taken into account in the intensity model, while the dust is possibly optically thick at this wavelength and the vertical structure may be relevant. However, as still more than 85\% of the structure is recovered by the Gaussian intensity equation and the SNR is very high, the description is sufficient to compare the azimuthal width at the two wavelengths. 

The residual of the 34 GHz image clearly shows a point source at the stellar position, with a peak flux of 36 $\mu$Jy beam$^{-1}$ (10$\sigma$). This point source can be either dust emission from an unresolved inner disk, or free-free/synchrotron emission from ionized gas close to the star. Since this emission is not seen in the ALMA data, the spectral index of  this emission is $<$1.3, so dust emission is unlikely. The origin of this emission can be determined using longer wavelength observations \citep{Rodmann2006,Ubach2012}. The total flux of the dust emission in the dust trap is thus only 216 $\mu$Jy at 34 GHz. 

\section{Discussion and summary}
The difference in azimuthal width between the two continuum images suggests a segregation of particle sizes, where centimeter-sized grains are more azimuthally  concentrated than millimeter-sized dust grains. However, optical depth effects can hide a narrow concentration of millimeter-dust grains equivalent to the 34 GHz morphology on top of the apparent millimeter-dust distribution. The optical depth $\tau_{680GHz}$ cannot be measured independently from the temperature with the available data. The measured brightness temperature at 680 GHz of 31 K (assuming the Planck equation) is only a factor of two lower than the calculated physical dust temperature of 60 K (at 60 AU radius in a physical disk model, \citealt{Bruderer2014}) so the emission is likely not highly optically thick. In order to quantify the optical depth at 680 GHz, we assume that the peak millimeter emission originates from an isothermal area $\Omega_{\rm beam}$ (sr) of the beam size with $T_{\rm dust}$=60 K. This is a conservative limit since $\tau$ likely decreases azimuthally outward. 
For the peak flux of 189 mJy, $\tau_{680GHz}$=0.5, so the emission is indeed marginally optically thick at 680 GHz. If the dust temperature is much lower than assumed here, the emission becomes fully optically thick and a narrow concentration of millimeter-dust grains could remain hidden. Locally lowering the temperature is possible by shadowing due to an inclined inner disk, such as proposed for HD142527 \citep{Marino2015} to explain the scattered light emission. For Oph IRS 48, there is no evidence for a local temperature drop, so we propose that the difference in emission originates from a spatial segregation between particles, where the centimeter-sized grains are more concentrated than the millimeter-sized grains. Spatially resolved continuum observations at intermediate wavelengths are required to confirm the dust temperature and optical depth. 


Combining the total 34 GHz flux of 216$\pm$22 $\mu$Jy with SMA and disk integrated ALMA observations at 230, 345 and 680 GHz gives a spectral slope of $\alpha = 2.84 \pm 0.06 (F_{\nu}\sim\nu^{\alpha}$). The fluxes are 50$\pm$7.5, 160$\pm$24 and 1000$\pm$200 mJy respectively, taken from \citet{Brown2012b} and this work, with the errors dominated by the flux calibration uncertainty. Measuring $\alpha$ independently between 680 and 230 GHz (2.8$\pm$0.2) and between 230 and 34 GHz (2.85$\pm$0.7) results in the same value within error bars, supporting at most modest optical depth at 680 GHz. 
 
\begin{figure}[!ht]
\begin{center}
\epsscale{1.}
\plotone{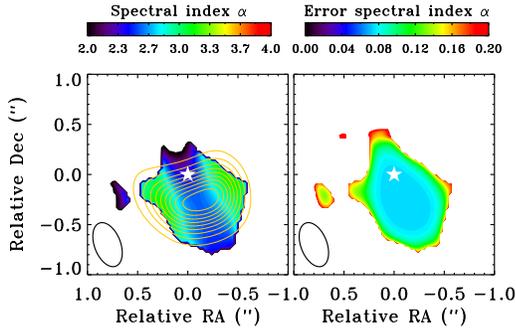}
\end{center}
\caption{Comparison of the ALMA and VLA data of Oph IRS 48, imaged with the same synthesized beam of 0$\farcs$46$\times$0$\farcs$26 (lower left corner). The left plot shows the spatially resolved spectral index $\alpha$, using 34 GHz and 680 GHz data. The right plot shows the error on $\alpha$ as function of position. The ALMA data are overlaid in yellow contours with 10\%,20\%,...,90\% of the peak flux. The spectral index decreases at the peak of the dust emission. Data points where the 34 GHz flux density is less than 2$\sigma$ ($\sigma=3.5\mu$Jy) are excluded. 
\label{fig:betaplot}}
\end{figure} 
 
Because both the VLA and the ALMA data are spatially resolved, the spectral index can be computed as a function of position (see Figure \ref{fig:betaplot}). For this figure, the ALMA data were convolved with the same synthesized beam as the VLA observations. The figure shows that $\alpha$ increases from 2.6$\pm$0.08 in the center to 3.3$\pm$0.15 in the outer wings, similar to the synthetic spectral index map from \citet{Birnstiel2013} of azimuthal trapping. The error in spectral index across the image is calculated based on the signal-to-noise ratio, as shown in the right panel of Figure \ref{fig:betaplot}. Therefore, the decrease of the spectral index at the peak of the dust emission is significant. Due to the flux calibration uncertainty, which is 20\% for ALMA Band 9 and 10\% for the VLA Ka band observations, the calibration error over the entire image on $\sigma_{\alpha}$ is $\sim0.07$, but this does not vary with position and therefore it is not included in the error plot.
 
Analysis of the dust opacity index $\beta$ ($\tau_{\rm \nu}\propto\nu^{\beta}$) is an important tool in constraining the dust particle sizes in the disk. When the dust emission is optically thin ($\tau\ll1$) and in the Rayleigh-Jeans regime, $\beta=\alpha-2$, where $\beta<1$ indicates dust growth to larger than millimeter-sized particles \citep{Draine2006}. For IRS 48 the 680 GHz is not optically thin and these assumptions cannot be made. Therefore, the dust opacity index thus has to be derived from the optical depth $\tau_{\nu}$ itself. 

Using $\tau_{\nu}\propto\nu^{\beta}$, $\beta$ can be calculated as a function of position by calculating $\tau_{\nu}$ at both frequencies at different positions, assuming $T_{\rm dust}$=60 K and the flux filling the beam. We find $\beta\sim0.7\pm0.1$ in the center and $\beta\sim1.3\pm0.3$ in the outer parts for a uniform temperature. 
As the continuum peak is likely more optically thick in the center, the emission there traces higher vertical layers where the temperature is higher, implying an even lower $\beta$: for $T_{\rm dust}$=200 K in the center, $\beta=0.6\pm0.1$ according to the same calculation. The azimuthal trend in $\beta$ is thus consistent with increased dust growth (increase of $a_{\rm max}$) or an increase of larger particles in the center (increase of $a_{\rm min}$). On the other hand, if the temperature is in fact as low as 31 K, $\beta$ would increase in the center. In order to get a uniform $\beta$ along the entire azimuthal shape (which implies no change in dust growth or size segregation), $T_{\rm dust}$ needs to be as low as 24 K at the center, which is more than a factor of two lower than the derived midplane dust temperature \citep{Bruderer2014}, while the optically thick emission likely originates from higher vertical layers with even higher temperatures. Thus, within reasonable azimuthal variations of the temperature, the results hint at variation in $\beta$, though some uncertainties remain.

In summary, the centimeter emission provides further support for the dust trapping mechanism in the Oph IRS 48 disk: within the assumptions of the 680 GHz optical depth and the temperature field, the centimeter-sized grains appear to be more concentrated in azimuth than the millimeter-sized grains as predicted in analytical dust models of azimuthal pressure maxima \citep{Birnstiel2013,LyraLin2013}. Inside the dust trap, grains may have reached centimeter sizes (and perhaps even larger), the first step toward planetesimal and planet formation.

\acknowledgments
We are grateful to E. F. van Dishoeck and M. Schmalzl for useful discussions, to the referee for useful comments and to the VLA staff for their efforts to observe this program after the configuration move. N.M. is supported by the Netherlands Research School for Astronomy (NOVA). J.T. acknowledges support from grant 639.041.439 from the Netherlands
Organisation for Scientific Research (NWO). T.B. acknowledges support from NASA Origins of Solar Systems grant NNX12AJ04G. Astrochemistry in Leiden is supported by the Netherlands
  Research School for Astronomy (NOVA), by a Royal Netherlands Academy
  of Arts and Sciences (KNAW) professor prize, and by the European
  Union A-ERC grant 291141 CHEMPLAN. The National Radio Astronomy Observatory is a facility of the National Science Foundation operated under cooperative agreement by Associated Universities, Inc. This paper makes use of the
  following ALMA data: ADS/JAO.ALMA/2013.1.00100.S. ALMA is a partnership of ESO (representing its member states), NSF (USA) and
  NINS (Japan), together with NRC (Canada) and NSC and ASIAA (Taiwan),
  in cooperation with the Republic of Chile. The Joint ALMA
  Observatory is operated by ESO, AUI/NRAO and NAOJ. 

{\it Facilities:} \facility{ALMA}, \facility{EVLA}.




\bibliographystyle{apj}

\end{document}